\documentclass[prl,twocolumn,amssymb,showpacs,floatfix,%
superscriptaddress]{revtex4}

\usepackage{graphicx}

\newcommand\apjl{Astrophys.~J. Lett.~}
\newcommand\apjs{Astrophys.~J. Suppl.~}
\newcommand\aap{Astron. Astrophys.~}

\begin{document}
%revtex3 \draft
%revtex3 \twocolumn[\columnwidth\textwidth\csname@twocolumnfalse\endcsname
\title{Improved models of stellar core collapse and still no
explosions:\\ What is missing?}
\author{R. Buras, M. Rampp, H.-Th. Janka, and K. Kifonidis}

\affiliation{Max-Planck-Institut f\"ur Astrophysik,
%revtex3 \address{Max-Planck-Institut f\"ur Astrophysik,
         Karl-Schwarzschild-Str.\ 1, D-85741 Garching, Germany}

\date{\today}
%revtex3 \maketitle

\begin{abstract}
Two-dimensional hydrodynamic simulations of stellar core-collapse
with and without rotation are presented which for the first time
were performed by solving the Boltzmann equation for the neutrino
transport including a state-of-the-art description of neutrino 
interactions.
Although convection develops below the neutrinosphere and in
the neutrino-heated region behind the supernova shock, the models
do not explode.
This suggests missing physics, possibly
with respect to the nuclear equation of state and 
weak interactions in the subnuclear regime. However, it might 
also indicate a fundamental problem of the neutrino-driven 
explosion mechanism.
\end{abstract}

%revtex3 ]

\maketitle

Despite of more than three decades of theoretical research 
and numerical modeling, the processes which cause the 
explosion of massive stars are still not understood.
Current observational data of supernovae (SNe) do not provide
direct information.
%  e.g. via radioactive nuclei that are 
%  created near the neutron star or asymmetries of the stellar 
%  ejecta. These have to be interpreted with great caution
%  concerning their meaning for the mechanism which starts the
%  explosion. 
Although neutrinos ($\nu$) and gravitational waves could yield
such insight, the $\nu$ events detected in connection
with SN~1987A were too sparse to constrain the 
SN mechanism. Progress in our understanding of the 
complex phenomena in collapsing stars and nascent (``proto-'') 
neutron stars (PNSs) is therefore mainly based on hydrodynamic
simulations.

Stars more massive than about 10 solar masses (M$_{\odot}$) 
develop an iron core in a sequence of nuclear burning stages.
This iron core becomes gravitationally unstable 
when it reaches a mass close to 
its Chandrasekhar limit and collapses to a neutron star.
A hydrodynamic shock forms when nuclear density is reached
and the matter becomes incompressible. There is general
agreement, supported by detailed numerical models, that this
shock is not able to promptly cause a SN explosion.
Instead, it
suffers from severe energy losses by the photodisintegration
of iron nuclei to free nucleons. It finally stalls after having
reached densities low enough that electron neutrinos ($\nu_e$) 
can rapidly
escape in a luminous outburst and thus drain even more energy from
the shock-heated matter~\cite{bru85,bru89ab,myra}. 

While $\nu$ losses damp the shock in this early phase,
the situation changes some 50$\,$ms later.
As more stellar matter falls onto the collapsed 
inner core, the shock is pushed to larger radii and the 
density and temperature behind the shock decrease. On the other
hand, the central core begins to settle and heats up, thus 
radiating more energetic neutrinos. Both effects 
lead to the situation that $\nu_e$ and $\bar\nu_e$
can now be absorbed with a small probability (10--20\%)
by free neutrons and protons behind the shock. A region of 
$\nu$ heating between the so-called ``gain radius'' and the 
shock front develops~\cite{betwil85}. If the $\nu$ energy
deposition is efficient enough, the stalled shock can be
revived and drives a ``delayed'' explosion.

The success of pioneering calculations~\cite{betwil85}
could be maintained in later models only by
invoking neutron-finger convective instabilities in the
PNS~\cite{wilmay8893}. These boost the $\nu$ luminosities
and thus enhance the $\nu$-energy transfer to the shock.
Explosion energies similar to those of observed SNe
required even higher $\nu$ fluxes. It was proposed that these
could be obtained when pions appear in large numbers in the
PNS matter~\cite{maytav93}. 
The existence of neutron-finger instabilities,
however, depends on very specific thermodynamical properties of the 
equation of state and on the details of the $\nu$
transport~\cite{brudin96}. 
%  It is also not clear whether
%  they can contribute to the energy transport on a macroscopic 
%  length scale and on a relevant timescale. 
The formation of 
pions in hot PNS matter, on the other hand, is highly
uncertain and requires particular assumptions about their 
dispersion relation.

While all simulations addressed so far were performed in 
one dimension (1D) assuming spherical symmetry (neutron-finger
convection was treated by a mixing-length approach), 
SN~1987A provided evidence for large-scale mixing
processes which carried radioactive nuclei from the region
of their formation near the PNS into the helium and
hydrogen shells of the exploding star. Simulations suggested
that their origin may be linked to hydrodynamic instabilities
behind the stalled shock already during the first second of the
explosion~\cite{herben92}. Two-dimensional (2D)~\cite{explconv}
and most recently also three-dimensional (3D)~\cite{frywar02} 
models that take into account $\nu$ effects then showed that
convective overturn indeed develops in the $\nu$ heating region 
and is helpful for shock revival, thus making explosions possible
even when spherically symmetric models fail~\cite{janmue96}. 

The multi-dimensional situation is generically different
because it allows accretion to continue while shock expansion
already sets in. Narrow downflows bring cold, low-entropy matter 
close to the gain radius, where the $\nu$ energy deposition 
is strongest. At the same time
heated matter can rise in buoyant bubbles, thus pushing 
the shock farther out and reducing energy loss by the
reemission of neutrinos. Although this increases the efficiency
of $\nu$-energy transfer, convection is still no
guarantee that explosions occur~\cite{janmue96}.
A particular concern with all multi-dimensional models which
yielded explosions was the much simplified treatment of the 
$\nu$ transport by grey diffusion schemes, which was inferior
to the more elaborate multi-group transport 
description used in unsuccessful spherical 
models~\cite{mezcal98:ndconv}. 

Recently it has become possible to go a step further
and solve the time-dependent Boltzmann equation 
for $\nu$ transport in 1D hydrodynamic simulations 
with Newtonian~\cite{ramjan00,mezlie01,thoetal02}  
and relativistic gravity~\cite{liemez01}. It turned out that
even this improvement 
does not lead to explosions. The question, however, remained
whether convection might bring the models to a success.
Here we present the first 2D simulations which
were performed with a Boltzmann solver for the $\nu$
transport. At the same time we have also upgraded the description 
of $\nu$-matter interactions compared to the conventional
treatment of Refs.~\cite{bru85,brumez93,ramjan02}.

{\em Numerical techniques and input physics.}
For the integration of the equations of hydrodynamics we employ
the Newtonian finite-volume code PROMETHEUS~\cite{frymue89}.
This second-order, time-explicit Godunov scheme 
is a direct Eulerian implementation of the
Piecewise Pa\-ra\-bol\-ic Method (PPM)~\cite{colwoo84} and
is based on a Riemann solver.

The Boltzmann solver scheme for $\nu$ and $\bar\nu$ of
all three flavors is described in detail in Ref.~\cite{ramjan02}. 
In multi-dimensional simulations in
spherical coordinates, we solve for each latitude $\theta$ of
the numerical grid the monochromatic moment equations for the
radial transport of $\nu$ number, energy, and momentum. This
set of equations is closed by a variable Eddington factor that is 
calculated from the solution of the Boltzmann equation 
on an angularly averaged stellar background. In
addition, we had to go an important step beyond this 
``ray-by-ray'' approximation of multi-dimensional transport. 
Physical constraints, namely the conservation of lepton
number and entropy within adiabatically moving fluid elements and
the stability of regions
which should not develop convection according to a stability
analysis, made it necessary to take into account the coupling of
neighboring rays at least by lateral advection terms and $\nu$
pressure gradients~\cite{buretal03}.

General relativistic effects are treated
approximately by modifying the gravitational potential with
correction terms due to pressure and energy of the stellar
medium and neutrinos, which
are deduced from a comparison of the Newtonian and relativistic
equations of motion~\cite{ramjan02}.
The $\nu$ transport contains gravitational
redshift and time dilation, but ignores the distinction between
coordinate radius and proper radius. This is
necessary for coupling the transport code to our basically 
Newtonian hydrodynamics. Comparison with fully relativistic
1D simulations showed that these approximations
work well at least when the deviations of the metric coefficients
from unity are moderate~\cite{lieetal03}.

Improving the description of $\nu$ interactions~\cite{buretal03} 
compared to the ``standard'' opacities~\cite{bru85,brumez93} we
added $\nu$-pair creation (and annihilation)
by nucleon-nucleon bremsstrahlung~\cite{hanraf98},
scattering of $\nu_{\mu}$, $\bar\nu_{\mu}$, $\nu_{\tau}$, 
and $\bar\nu_{\tau}$
off $\nu_e$ and $\bar\nu_e$, and pair annihilation reactions
between neutrinos of different flavors~\cite{burjan02:nunu}.
We also take into account the detailed reaction
kinematics with nucleon thermal motions, recoil, and
fermion phase-space blocking effects in the
charged- and neutral-current $\nu$-nucleon interactions.
Moreover, nuclear correlations \cite{bursaw9899}, weak-magnetism
corrections~\cite{hor02}, and the 
reduction of the nucleon effective mass and possible
quenching of the axial-vector coupling in nuclear
matter~\cite{carpra02} are included, too.

\begin{table}[htb]
\caption{Parameters of computed 2D models for
progenitor stars with different masses. $M_{\mathrm{Fe}}$
is the iron core mass, $M_{\mathrm{Si+O}}$ the mass 
interior to the inner edge of the oxygen-rich Si-shell,
$\Omega_{\mathrm{i}}$ the angular velocity of
the Fe-core prior to collapse, $\theta_0$ and
$\theta_1$ are the polar angles of the lateral grid boundaries,
and $N_{\theta}$ is the number of grid points in 
lateral direction.}
\setlength\tabcolsep{5pt} % enlarge column spacing
%revtex3 \setlength\tabcolsep{6pt} % enlarge column spacing
\begin{tabular}{lllllll}
\hline
\hline
Model & Mass          & $M_{\mathrm{Fe}}$ & $M_{\mathrm{Si+O}}$ 
      & $\Omega_{\mathrm{i}}$ & $[\theta_0,\theta_1]$ 
      & $N_{\theta}$  \\
      & (M$_{\odot}$) & (M$_{\odot}$) & (M$_{\odot}$) 
      & (s$^{-1}$)            & (degrees)            &     \\
\hline
s11.2  & 11.2 & 1.24 & 1.30 & 0   & $[46.8,133.2]$ & 32 \\
s11.2  & 11.2 & 1.24 & 1.30 & 0   & $[46.8,133.2]$ & 64 \\
s15    & 15   & 1.28 & 1.43 & 0   & $[46.8,133.2]$ & 32 \\
s15p   & 15   & 1.28 & 1.43 & 0   & $[46.8,133.2]$ & 64 \\
s15r   & 15   & 1.28 & 1.43 & 0.5 & $[0,90]$       & 64 \\
s20    & 20   & 1.46 & 1.46 & 0   & $[46.8,133.2]$ & 32 \\
\hline
\hline
\end{tabular}
\label{tab:models}
\end{table}

Our current SN models are calculated with the
equation of state (EoS) of Ref.~\cite{latswe91} using
an incompressibility of bulk nuclear matter of 180$\,$MeV 
(other values make only minor differences~\cite{thoetal02}).
This EoS is based 
on a compressible liquid-drop model 
including nuclei, nucleons, $e^-$, $e^+$, and photons.

The 2D simulations were performed with a spherical coordinate grid
with about 400 radial zones
and assuming azimuthal symmetry around the polar axis. A wedge of
roughly 90 degrees was chosen in the angular direction between the
angles $\theta_0$ and $\theta_1$ and with the number of mesh points,
$N_{\theta}$, as given in Table~\ref{tab:models}. For the
nonrotating models periodic boundary conditions were taken.
For the rotating model reflecting boundaries at the axis
and the equator were imposed.

We considered three (solar metallicity) progenitors with 
main sequence masses of 11.2$\,$M$_{\odot}$, 15$\,$M$_{\odot}$, 
and 20$\,$M$_{\odot}$~\cite{wooheg02}. The model specific 
parameters are given in Table~\ref{tab:models}. For the
transport we used an energy grid of 17 geometrically spaced
bins with centers from 2$\,$MeV to 333$\,$MeV. Higher energy
resolution was tested without giving significant differences.
Model~s15p is
a calculation where random density perturbations with 1\% amplitude
were imposed on the pre-collapse core to follow their growth
during infall~\cite{laigol00}. In the rotating Model
s15r the angular frequency was assumed to be constant in the 
Fe and Si core and decreasing like $r^{-3/2}$ outside of
1750$\,$km (1.43$\,$M$_{\odot}$). These choices and the adopted
rotation rate (cf.~Table~\ref{tab:models}) are basically in agreement
with predictions from stellar evolution models~\cite{hegwoo03}.

\begin{figure}[ht]
  \begin{center} \leavevmode
   \includegraphics[width=0.99\columnwidth]{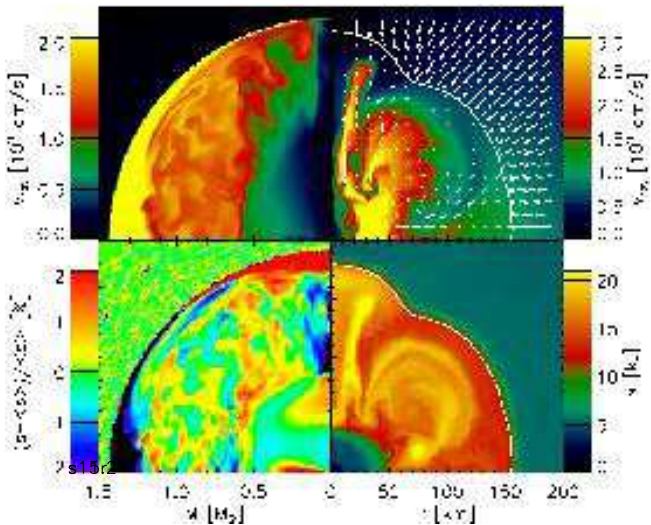}
   \vspace{0.25cm}
    \caption{Snapshots of the stellar structure for the rotating
         Model~s15r at 198~ms after shock formation. The left 
         panels show the rotational velocity (top) and the
         fluctuations of entropy (in per cent) versus the 
         enclosed mass, emphasizing the conditions inside the
         neutron star. The right panels display the rotational
         velocity (top) and the entropy (in $k_{\mathrm{B}}$ per
         nucleon) as functions of radius. The arrows indicate 
         the velocity field, the white line marks the shock front.
     \label{fig:rotmod}}
   \end{center}   
\end{figure}

\begin{figure}[ht]
  \begin{center} \leavevmode
   \includegraphics[width=0.99\columnwidth]{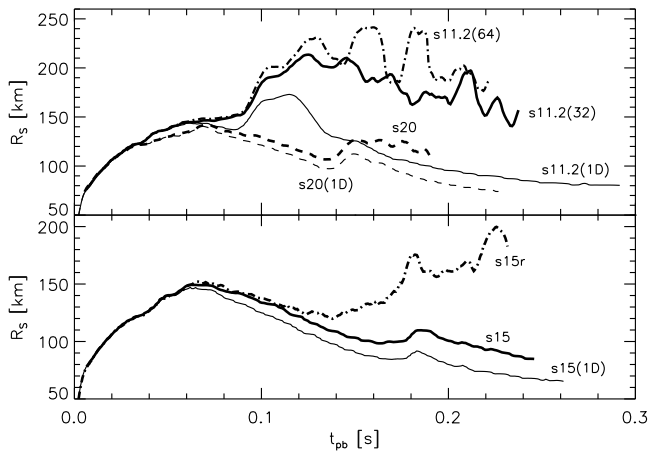}
   \vspace{0.25cm}
    \caption{Shock radii $R_{\mathrm{s}}$ (averaged over polar angles) 
             vs. post-bounce time $t_{\mathrm{pb}}$. The 2D
             models (bold lines) are compared to the corresponding
             1D simulations (thin lines).
     \label{fig:rshock}}
   \end{center}
\end{figure}

\begin{figure}[ht]
  \begin{center} \leavevmode
   \includegraphics[width=0.99\columnwidth]{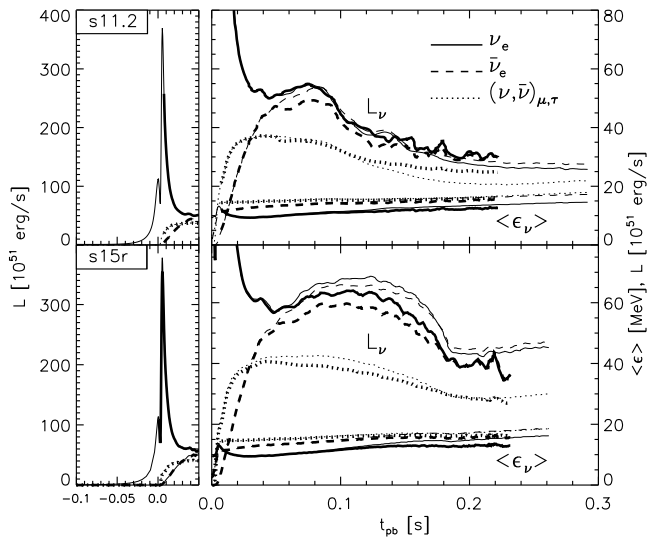}
   \vspace{0.25cm}  
    \caption{Luminosities and mean energies
             (defined as ratio of energy flux to number flux) for $\nu_e$,
             $\bar\nu_e$ and $\nu_{\mu,\tau}$, $\bar\nu_{\mu,\tau}$
             vs.~time for Models~s11.2 (top) and s15r (measured
             by an observer 
             comoving with the stellar medium at 500$\,$km). 
             The left panels (left scale) show
             the prompt $\nu_e$ burst, the right panels
             enlarge the post-bounce evolution. The thin lines represent
             results of 1D simulations for comparison.
     \label{fig:neutrinos}}
   \end{center}
\end{figure}

{\em Results.} Convective overturn starts to appear in the 
region of $\nu$ heating behind the shock about 25$\,$ms after
shock formation, and begins to affect the SN dynamics 
about 50$\,$ms later.
Expanding bubbles of heated matter deform the shock 
(Fig.~\ref{fig:rotmod}) and lead to a transient increase of
the average shock radius (Fig.~\ref{fig:rshock}). The 
difference to spherically symmetric simulations, however,
is small in case of Models~s15 and s20 because the convectively
unstable layer between gain radius and shock remains narrow and
the overturn motions never become very strong. The largest
relative change occurs when the big entropy discontinuity
at the inner boundary of the oxygen enriched shell 
($M_{\mathrm{Si+O}}$; Table~\ref{tab:models}) of
the progenitor star crosses the shock. The sudden decrease
of the density and ram pressure of the infalling matter 
allows the shock to transiently expand also
in 1D simulations. In Model~s11.2 the corresponding
growth of the shock radius is nearly 50$\,$km. This in turn
widens the gain layer and thus allows convection to strengthen.
The combined effects lead to a much larger shock radius than
in the 1D case (Fig.~\ref{fig:rshock}). 
A similar result is caused by centrifugal forces in the
rotating Model~s15r (Fig.~\ref{fig:rshock}). Here decelerated
infall of matter near the equatorial plane and a reduced 
density near the poles create a more favorable situation for
violent convection (Fig.~\ref{fig:rotmod}).

Ledoux convection sets in {\em below} the
neutrinosphere already about 40$\,$ms after bounce.
%  (Fig.~\ref{fig:nsconvection}). 
It is persistent until the end of our simulations and 
slowly digs farther into the star 
(cf.~Fig.~\ref{fig:rotmod}, lower left panel) 
in agreement with previous
2D simulations~\cite{keietal96}. But the active layer
is rather deep inside the PNS (at a density above 
$10^{12}\,$g$\,$cm$^{-3}$) and is surrounded by a convectively
stable shell in which the surface fluxes of $\nu_e$ 
and $\bar\nu_e$ are mainly built up. Therefore the PNS convection 
has little influence on the
emission of these neutrinos and is irrelevant for the
SN dynamics. The differences in the 
$\nu_e$ and $\bar\nu_e$ luminosities and mean energies relative 
to the 1D case (Fig.~\ref{fig:neutrinos}) are caused by
rotation and 
matter downflows from the shock to the neutrinospheric region.

Random density perturbations in the progenitor (Model s15p) increase
in the supersonically infalling layers and are damped in the
subsonically collapsing inner core as predicted by linear
analysis~\cite{laigol00}. We have not discovered any
influence on the development of convection. Also the
shock radii of Models~s15p and s15 are nearly identical.
Higher angular resolution allows convection to grow somewhat
faster because of reduced numerical viscosity. Within the
limits of our tests the later evolution in cases of strong
convection exhibits quantitative but no qualitative differences
(see Models s11.2 in Fig.~\ref{fig:rshock}).

{\em Conclusions.}
Our simulations show that an accurate
treatment of the neutrino physics does not yield sufficiently
efficient $\nu$-energy transfer behind the stalled SN shock
to produce explosions, even though convection occurs below the
neutrinosphere and in the $\nu$-heating region.
This finding confirms concerns~\cite{mezcal98:ndconv}
that the success of previous models~\cite{explconv} was
favored by gross simplifications of the $\nu$ transport.

% We have now reached a level of refinement that our main
% conclusions should not be affected by remaining numerical 
% deficiencies. The
% absence of explosions thus points to deficits in the 
% input physics. 
None of the included effects is therefore able to cause
explosions. 
Truly multi-dimensional transport~\cite{buretal03} or full 
relativity~\cite{lieetal03} are not likely to change the situation.
% This points to deficits in the input physics.
But the long-time evolution
of the shock is sensitive to the $\nu$ emission that
originates from the neutrinospheric layer. The EoS properties 
and $\nu$ interactions at densities below some
$10^{13}\,$g$\,$cm$^{-3}$ are therefore particularly relevant.
The structure, temperature, and convective stability of this 
layer depend also
on the compactness of the PNS and thus on the 
uncertain physics in its supranuclear core. The influence of
the EoS on the post-bounce
dynamics, however, has not been explored extensively so far.
Necessary improvements of weak interactions on nuclei include 
electron capture rates~\cite{lanetal03} and inelastic neutrino
scattering. Moreover, generically 3D phenomena cannot be studied 
with 2D hydrodynamics. Also the properties of the 
progenitor cores need reexamination.

The current paradigm for explaining massive star explosions
would have to be revised if the $\nu$-driven mechanism were
fundamentally flawed.
SNe might be aided by, e.g., magnetohydrodynamic
processes~\cite{akietal02} or even more exotic physics.
Although it is still too early for this conclusion,
such investigations deserve more interest.

We are indebted to K.~Takahashi and C.~Horowitz for
help in improving $\nu$-nucleon interactions.
Support by the Sonderforschungsbereich
375 on ``Astroparticle Physics'' of the Deutsche 
Forschungsgemeinschaft is acknowledged.
The simulations were done on the IBM ``Regatta''
supercomputer of the Rechenzentrum Garching.


\begin{thebibliography}{99}

\bibitem{akietal02}
S. Akiyama, J.C. Wheeler, D.L. Meier, and I. Lichtenstadt,
\apj, in press ({\tt astro-ph/\-0208128})

\bibitem{betwil85}
J.R. Wilson, in \emph{Numerical Astrophysics}, edited by J.M.~Centrella,
{\em et al.}
% J.M.~LeBlanc, R.L.~Bowers, and J.A.~Wheeler
(Jones and Bartlett, Boston, 1985) 422;
H.A. Bethe and J.R. Wilson, \apj {\bf 295}, 14 (1985)

\bibitem{bru85}
S.W. Bruenn, \apjs {\bf 58}, 771 (1985)

\bibitem{bru89ab}
S.W. Bruenn, \apj {\bf 340}, 955 (1989); 
\apj {\bf 341}, 385 (1989)

\bibitem{brudin96}
S.W. Bruenn and T. Dineva, \apjl {\bf 458}, L71 (1996)

\bibitem{burjan02:nunu}
R. Buras, H.-T. Janka, M.-T. Keil, G. Raffelt, and M. Rampp,
\apj, in press ({\tt astro-ph/\-0205006})

\bibitem{buretal03}
R. Buras, M. Rampp, H.-T. Janka, K. Kifonidis, K. Takahashi, and
C.J. Horowitz, in preparation

\bibitem{bursaw9899}
A. Burrows and R.F. Sawyer, Phys. Rev. C. {\bf 58}, 554 (1998);
Phys. Rev. C. {\bf 59}, 510 (1999)

\bibitem{explconv}
A. Burrows, J. Hayes, and B.A. Fryxell, \apj {\bf 450}, 830 (1995);
M. Herant, W. Benz, W.R. Hix, C.L. Fryer, and S.A. Colgate,
\apj {\bf 435}, 339 (1994);
C.L. Fryer, \apj {\bf 522}, 413 (1999);
C.L. Fryer and A. Heger, \apj {\bf 541}, 1033 (2000)

\bibitem{carpra02}
G.W. Carter and M. Prakash, Phys. Lett. B {\bf 525}, 249 (2002)

\bibitem{colwoo84}
P. Colella and P.R. Woodward, 1984, J.~Comp. Phys. {\bf 54}, 174

\bibitem{frywar02}
C.L. Fryer and M.S. Warren, \apjl {\bf 574}, L65 (2002)

\bibitem{frymue89}
B.A. Fryxell, E. M\"uller, and W.D. Arnett, 
Preprint MPA-449 (MPI f\"ur Astrophysik, Garching,
1989)

\bibitem{hanraf98}
S. Hannestad and G. Raffelt, \apj {\bf 507}, 339 (1998)

\bibitem{hegwoo03}
A. Heger, S.E. Woosley, and H. Spruit, \apj, in preparation

\bibitem{herben92}
M. Herant, W. Benz, and S.A. Colgate, \apj {\bf 395}, 642 (1992)

\bibitem{hor02}
C.J. Horowitz, Phys. Rev. D {\bf 65}, 043001-1 (2002)

\bibitem{laigol00}
D. Lai and P. Goldreich, \apj {\bf 535}, 402 (2000)

\bibitem{janmue96}
H.-T. Janka and E. M\"uller, \aap {\bf 306}, 167 (1996)

\bibitem{keietal96}
W. Keil, H.-T. Janka, and E. M\"uller, \apjl {\bf 473}, L111 (1996)

\bibitem{lanetal03}
K. Langanke, {\em et al.}, \prl, submitted (2003)

\bibitem{latswe91}
J.M. Lattimer and F.D. Swesty, Nucl. Phys. {\bf A535}, 331 (1991);
J.M. Lattimer, C.J. Pethick, D.G. Ravenhall, and D.Q. Lamb,
Nucl. Phys. {\bf A432}, 646 (1985)

\bibitem{lieetal03}
M. Liebend\"orfer, M. Rampp, H.-T. Janka, and A. Mezzacappa, 
in preparation

\bibitem{liemez01}
M. Liebend\"orfer, {\em et al.},
% A. Mezzacappa, F.-K. Thielemann, O.E.B. Messer, 
% W.R. Hix, and S.W. Bruenn, 
Phys. Rev. D. {\bf 63}, 3004 (2001)

\bibitem{maytav93}
R.W. Mayle, M. Tavani, and J.R. Wilson, \apj {\bf 418}, 398 (1993)

\bibitem{brumez93}
A. Mezzacappa and S.W. Bruenn, \apj {\bf 405}, 637 (1993);
\apj {\bf 410}, 740 (1993)

\bibitem{mezlie01}
A. Mezzacappa, {\em et al.},
% M. Liebend\"orfer, O.E.B. Messer, W.R. Hix, 
% F.-K. Thielemann, and S.W. Bruenn, 
\prl {\bf 86}, 1935 (2001)

\bibitem{mezcal98:ndconv}
A. Mezzacappa, {\em et al.},
% A.C. Calder, S.W. Bruenn, J.M. Blondin, M.W. Guidry, 
% M.R. Strayer, and A.S. Umar, 
\apj {\bf 495}, 911 (1998)

\bibitem{myra}
E.S. Myra, {\em et al.},
% S.A. Bludman, Y. Hoffman, I. Lichtenstadt, N. Sack,
% and K.A. van Riper, 
\apj {\bf 318}, 744 (1987);
E.S. Myra and S.A. Bludman, \apj {\bf 340}, 384 (1989)

\bibitem{ramjan00}
M. Rampp and H.-T. Janka, 2000, \apjl {\bf 539}, L33 (2002)

\bibitem{ramjan02}
M. Rampp and H.-T. Janka, \aap {\bf 396}, 361 (2002)

\bibitem{thoetal02}
T.A. Thompson, A. Burrows, and P.A. Pinto, \apj, in press
({\tt astro-ph/\-0211194})

\bibitem{wilmay8893}
J.R. Wilson and R. Mayle, Phys. Rep. {\bf 163}, 63 (1988);
Phys. Rep. {\bf 227}, 97 (1993)

\bibitem{wooheg02}
S.E. Woosley, A. Heger, and T.A. Weaver, Rev. Mod. Phys. {\bf 74},
1015 (2002)


\end{thebibliography}
\end{document}